# The phase transition in VO2 probed using x-ray, visible and infrared radiations

Suhas Kumar[1], John Paul Strachan[1*], A. L. David Kilcoyne[2], Tolek Tyliszczak[2], Matthew D. Pickett[1], Charles Santori[1], Gary Gibson[1] and R. Stanley Williams[1]

[1] *Hewlett Packard Laboratories, 1501 Page Mill Road, Palo Alto, California, 94304, USA,*

[2] *Advanced Light source, Lawrence Berkeley National Laboratory, Berkeley, California, 94720, USA*

Email: John-Paul.Strachan@HP.com

*Vanadium dioxide (VO2) is a model system that has been used to understand closely-occurring multiband electronic (Mott) and structural (Peierls) transitions for over half a century due to continued scientific and technological interests. Among the many techniques used to study VO2, the most frequently used involve electromagnetic radiation as a probe. Understanding of the distinct physical information provided by different probing radiations is incomplete, mostly owing to the complicated nature of the phase transitions. Here we use transmission of spatially averaged infrared (λ=1.5 µm) and visible (λ=500 nm) radiations followed by spectroscopy and nanoscale imaging using x-rays (λ=2.25-2.38 nm) to probe the same VO2 sample while controlling the ambient temperature across its hysteretic phase transitions and monitoring its electrical resistance. We directly observed nanoscale puddles of distinct electronic and structural compositions during the transition. The two main results are that, during both heating and cooling, the transition of infrared and visible transmission occur at significantly lower temperatures than the Mott transition; and the electronic (Mott) transition occurs before the structural (Peierls) transition in temperature. We use our data to provide insights into possible microphysical origins of the different transition characteristics. We highlight that it is important to understand these effects because small changes in the nature of the probe can yield quantitatively, and even qualitatively, different results when applied to a non-trivial multiband phase transition. Our results guide more judicious use of probe type and interpretation of the resulting data.*

The two transitions in vanadium oxide have been a matter of curiosity and debate over several decades because of the apparently simultaneous electronic (Mott) and structural (Peierls) transitions and the rich physics involved.[1,2] There have been many studies on the electronic, optical and mechanical properties of VO2, owing to promising technological applications such as information storage and micromechanical actuation.[3-6] Some of the important scientific issues recently addressed include the nature and sequence of the transitions,[7,8] effect of surface charges,[9,10] role of joule heating during resistance switching,[11,12] phase diagram of the transitions,[13] etc. There have been debates and contradicting results on several of these issues.[12,14-16]

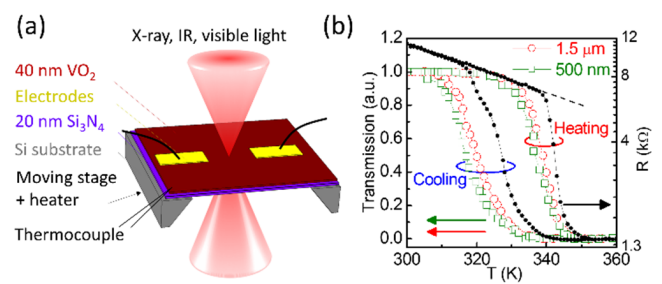

**Figure 1**: (a) Schematic of the setup. (b) Normalized transmission of infrared (1.5 µm) and visible light (500 nm) during heating and cooling (left axis), along with change in resistance on a common temperature axis. Black dashed line is a guide to the eye that follows the linear change in log(R) with temperature at low temperatures.

Many of these studies employ electromagnetic radiation as a probe of the underlying science. For example, the change in electrical resistance due to a Mott transition in $VO_2$ has been studied using x-ray photoemission spectroscopy,[17] visible light reflection,[15] and single-energy infrared transmission.[18,19] Here we present measurements on the same temperature-controlled thin film of $VO_2$ using infrared transmission (1.5 µm), visible light transmission (500 nm) and high-resolution spectroscopy and nanoscale imaging using x-ray (2.25-2.38 nm), while simultaneously monitoring its resistance. We are able to show that each has a distinctly different response. The important results are: (1) the critical transition temperature ($T_C$, defined later) of infrared and visible transmission is significantly lower than the electronic (Mott) critical transition temperature (i.e.: $T_C^{visible} < T_C^{infrared} < T_C^{electronic}$, during both heating and cooling) and, (2) the electronic (Mott) transition precedes the structural (Peierls) transition in temperature (i.e.: $T_C^{electronic} < T_C^{structural}$ during heating and $T_C^{structural} < T_C^{electronic}$ during cooling). Using this data set, we consider physical origins of these differences, including light scattering by sub-micrometer sized puddles of distinct phases. We also suggests directions for future studies to address questions raised by this work.

We grew 40 nm of $VO_2$ using a precursor oxidation process on 20 nm thick silicon nitride films suspended over holes etched into a silicon wafer that enabled transmission measurements at all wavelengths of interest.[20,21] In-plane Pt electrodes were deposited with a gap of 4 µm between them to enable simultaneous in-situ monitoring of low-bias film resistance (Figure 1a), which also served as a temperature calibration between different experiments. Ambient temperature was controlled (within ±0.1 K) using a stage heater with feedback of film temperature from a thermocouple. An infrared beam of wavelength 1.5 µm was directed onto the sample and the spatially averaged transmitted intensity was measured (Figure 1b). When heated, the normalized transmission stayed approximately constant up to about 335 K, then decreased until about 345 K, after which it flattened out at its minimum. The low-bias resistance of the film seemed to deviate from its low-temperature behavior at a relatively higher temperature of about 340 K during heating. Upon cooling the film, the well-known hysteresis was observed in both resistance and infrared transmission. During both heating and cooling, the temperature of transition of infrared transmission was lower than the temperature of transition of resistance. The transmission of 500 nm visible light followed a similar hysteretic transition and the temperature of the transition of visible light was further down-shifted relative to that of the infrared transmission during both heating and cooling. This temperature downshift has been previously observed,[22-24] but its origins have not been experimentally explored.

X-ray absorption spectromicroscopy maps of the oxygen K-edge (E=520-550 eV, $\lambda$=2.38-2.25 nm) were taken on the same sample, using the system for temperature control and low-bias resistance monitoring described above. Figures 2a-2b display a set of spatially averaged spectra taken at temperatures that span the heating and cooling transition temperatures. Two distinct changes in the spectra are the downshift of the $\pi^*$ band and the vanishing of the $d_\parallel^*$ band in the high temperature phase (rutile metal) relative to the low temperature phase (monoclinic insulator) (Figure 2f).[2] It was shown that the lowest conduction band ($\pi^*$) accounts for the conductivity of the material while the $d_\parallel^*$ band indicates the structural distortion; hence the Mott transition and the Peierls transitions can be separately tracked by tracking the evolution of the $\pi^*$ and the $d_\parallel^*$ bands, respectively.[15,25] The gradually changing spectra in the intermediate temperatures in Figures 2a-2b are averages over many coexisting nanoscale domains with two different conductivities and structures that spatially evolve during the transition.[8,26]

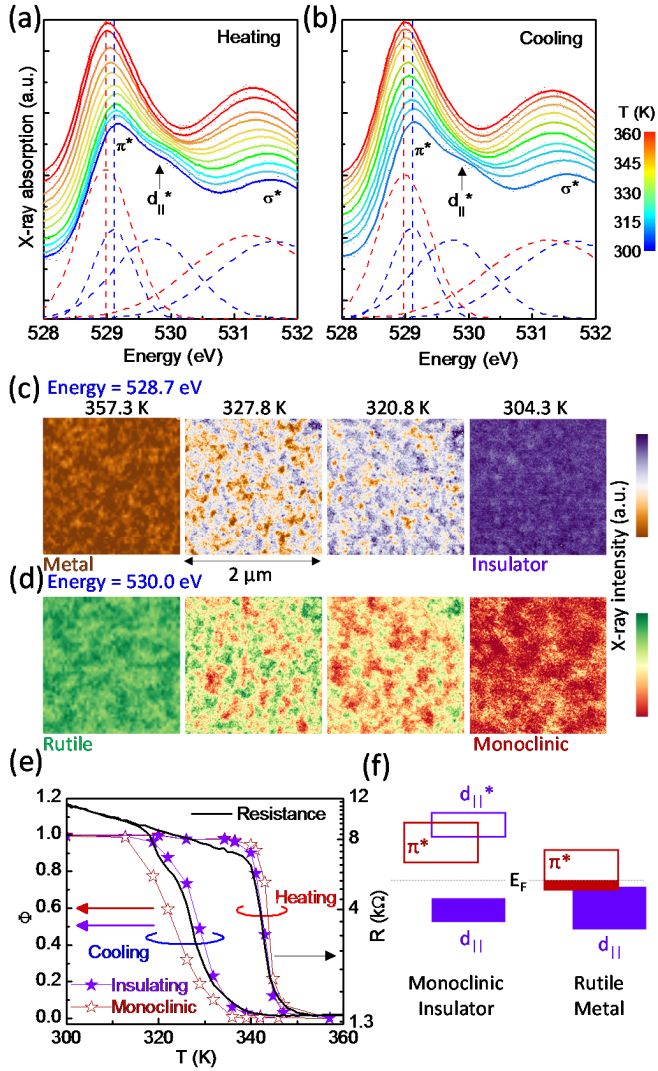

**Figure 2:** (a)-(b) X-ray oxygen K-edge spatially averaged spectra across the transition temperatures during both heating and cooling. The fits to the extreme spectra (color coded) are shown in each panel along with annotations of the energy bands. Vertical dashed lines represent the position of the lowest conduction band for the extreme spectra in each panel. (c)-(d) Spatially resolved x-ray maps at two different energies as noted, for four different temperatures across the transition during cooling. (e) Insulating and monoclinic phase fractions ($\Phi$) calculated from data in (a)-(b). Resistance data from Figure 1b is included for comparison. (f) Band schematics of the extreme phases.

transition.[8,26] Using focused, high intensity x-rays with spatial and spectral resolutions of ~25 nm and <70 meV, respectively,[27] we mapped an area of the film at two different x-ray energies – 528.7 eV, which corresponds to the $\pi^*$ band, to correlate to the electrical conductivity, and 530 eV, which corresponds to the $d_\parallel^*$ band, to chart the structural evolution (Figures 2c-2d). These maps reveal the spatial evolution of domains of different phases and provide an insight into the length scales involved. Because we have maps at only two temperature points during the transition and due to a low signal-noise ratio, we have not used this data to provide direct quantitative estimates of the phase fractions or evidence of intermediate states (see Supplemental Material, Section 5).[28] Thus, we calculated the phase fractions of the insulating and monoclinic components separately, using peak-fits to the spectra in Figures 2a-2b and observed the evolution of the $\pi^*$ and $d_\parallel^*$ bands. The main observation is that the electronic transition (insulator to metal) precedes the structural transition (monoclinic to rutile) in temperature, during both heating and cooling (Figure 2e). This is very consistent with our previous experimental result using a more direct and detailed mapping of the two transitions in the same film of $VO_2$ used here.[8]

To better compare the transitions observed in the infrared, visible and x-ray transmission, and the resistance, we used the 2-dimensional Bruggeman effective medium approximation (EMA) for percolative media to calculate the macroscopic resistance (black curve in Figure 3a) by parametrically varying the volume fraction of the component phases with different resistivities (using terminal resistance values from Figure 1b for heating and cooling braches separately).[29-31] We then use the normalized transmission of Figure 1b as indicative of the phase fractions ($\Phi$) measured by infrared and visible lights and map them to the resistance of the film at the corresponding temperatures (Figure 3a). As it will be clear in the following paragraphs, this supposition/assumption is made to explicitly discount optical transmission as a direct measure of phase fraction, as claimed before.[18] For example, a fraction of 0.5 of infrared transmission occurs at ~320 K, where the resistance of the film is ~7 kΩ. A similar mapping of the phase fractions in Figure 2e to the resistance of the film is also plotted in Figure 3a. This is a direct comparison of the different quantities involved on a common scale. From this plot and inspection of Figures 1b and 2e, it is apparent that, relative to the change in resistance, infrared transmission is downshifted in temperature across the entire hysteresis, with a larger width of the hysteresis. Visible light transmission is further downshifted in temperature. The change in insulating fraction appears to follow the change in resistance, but the change in structural composition is upshifted on the heating branch and downshifted on the cooling branch. To quantify this argument, we define a critical phase fraction, $\Phi_C$=0.5 (Figure 3a), which defines the critical temperature, $T_C$, for each of the transitions. Using Figure 1b, we map the corresponding resistance values to $T_C$ on both the heating and cooling branches, as reported in Figure 3b. The variation in $T_C$ for the different transitions reaffirms the observations presented above and also shows that the change in the insulating phase fraction is the closest match to the change in resistance, as one would intuitively expect. The purpose of Figure 3 is to quantitatively present the results that are fairly apparent from an inspection of Figures 1b and 2e.

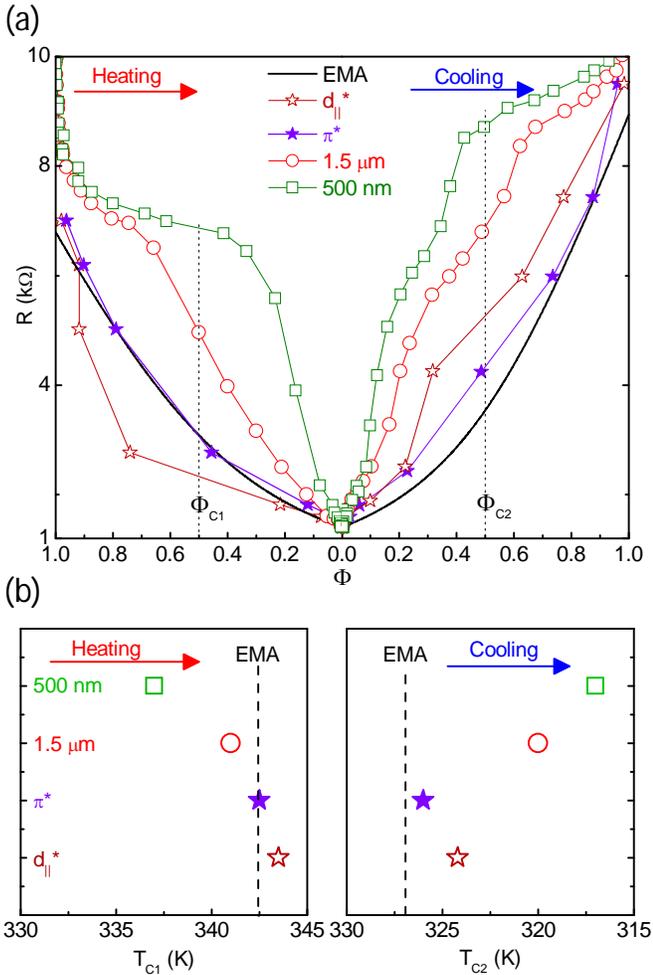

(a)

(b)

**Figure 3:** (a) Resistance plotted against phase fraction, $\Phi$, calculated for different measurements during heating and cooling. While resistivity and x-ray data are indeed volume fractions, transmission data are fractional transmission values. $\Phi_{C1}$ and $\Phi_{C2}$ are arbitrarily defined critical phase fractions (0.5) on the heating and cooling branches, for which critical temperatures, $T_{C1}$ and $T_{C2}$, are reported in (b). Y-axis is arbitrarily scaled and offset to indicate results from different measurements, as noted in the legend. Dashed vertical lines in (b) indicate $T_{C1}$ and $T_{C2}$ calculated from EMA during heating and cooling.

An intuitive possibility to explain the downshifting of infrared transmission noticed above is that the decrease in infrared transmission is due to Mie scattering by puddles of metallic regions developing during the Mott transition; while a change in resistance requires that the metallic puddles have a high enough density to form a conducting pathway, thereby occurring at a higher temperature.[24] From Figure 2c, we note that the metallic puddles are on the length scale of ~100 nm at the intermediate temperatures, whereas infrared transmission continues to change (Figure 1b, during cooling) even over temperatures for which the metallic puddles are expected to be ≤100 nm. This supports the supposition that the infrared transmission changes despite the fact that no conducting pathway exists, the latter being required to observe a change in resistance. But beyond that, given that the wavelength used was 1.5 μm, a set of sparsely distributed 100 nm ellipsoids are highly unlikely to cause significant Mie scattering or Rayleigh scattering (see Supplemental Material, Section 4, for calculations).[28] Since the corresponding length scales in structural inhomogeneity (Figure 2d) are relatively larger (200-300 nm), and closer to the infrared wavelength, one might believe that it could cause the transition in the infrared transmission through scattering. But the structural transition follows a qualitatively different transition path compared to the infrared transmission, recalling the $T_C$ relationships previously mentioned (Figures 3a-3b), so the structural inhomogeneity is unlikely to have primarily caused scattering-induced change in infrared transmission. There are two recently discovered intermediate phases during the transition of $VO_2$, one each on the heating and cooling branches (i.e.: monoclinic metal during heating and rutile insulator during cooling), which appear only within a narrow range of transition temperatures under certain experimental and film-growth conditions.[8,32,33] Because they contribute to the spatial inhomogeneity, it is necessary to account for their optical behavior to explain the significant temperature shifts in the transitions of infrared and visible light transmission. In order to understand this, it is necessary to measure their optical constants, which would require stabilizing nanoscale puddles of them during the transition while performing optical measurements, potentially using a near-field visible-infrared high-resolution spectromicroscopy.[7,26,34] The challenging nature of this experiment is probably the reason the intermediate phases have never been spectrally isolated using optical techniques. An example of an approach to isolating the intermediate states using local joule heating is presented in Figure 4 and experimentally detailed elsewhere.[11,28]

The electronic transition preceding the structural transition during both heating and cooling has been observed before and also theoretically modeled,[8,32,35-37] and due to the sensitivity of the $VO_2$ system to strain, doping, etc., the transitions have been observed to have a different sequence and/or intermediate states depending on growth and experimental conditions.[2,15-17,38,39] Here we merely highlight that high-resolution spectroscopy is essential to directly isolate the two transitions and the possible intermediate states involved,[7,8] while an arbitrary choice of the probing wavelength, as is sometimes used in the infrared and visible spectral range,[18,19] can provide misleading or no information on the phase transition. To further emphasize this, we performed temperature dependent visible light reflection experiments to show that certain wavelengths yield no information on the existence of a phase transition (see Supplemental Material, Figure S1).[28] Some of the reasons why low-resolution optical measurements face challenges in the study of the $VO_2$ phase transitions are: 1) There is sub-wavelength spatial inhomogeneity during the transition, thereby requiring a near-field nanoscale spatial resolution, 2) The bandgap is small (<1 eV) and the band edges are smeared out in energy, thereby often preventing a study of the evolution of the individual energy bands during the phase transition,[40] 3) Multiple optical transitions occur through evolution of multiple bands overlapping in energy ($\pi^*$ and $d_{\parallel}^*$), thereby requiring a high spectral resolution.

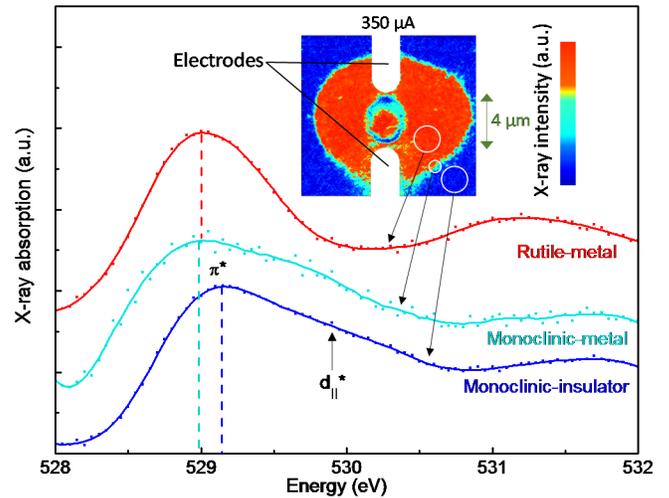

**Figure 4**: X-ray transmission map (inset) of the same device as the rest of this study with a large current forced between the electrodes (throughout this study, a small voltage (<0.1 V) was used to measure low-bias resistance). The differently colored region (or filament) between the electrodes indicates the region that underwent a phase transition due to high temperatures induced by joule heating. Spectra corresponding to the three different regions, as indicated (main panel). Inside and outside the filament, we find the extreme phases of $VO_2$ that are well known, namely the high-temperature rutile metal and the low-temperature monoclinic insulator, respectively. At the edge of the filament, where the temperatures are expected to be in the vicinity of the transition temperature, we find a new phase which has the signatures of a metal (with $\pi^*$ band at a lower energy) with a monoclinic phase ordering (with the significant presence of the $d_{\parallel}^*$ band). This intermediate phase is expected because the current was increased from 0 to 350 μA and held constant at 350 μA while the x-ray spectromicrograph was obtained. Hence, the sample was heated (and not cooled), thereby justifying the presence of the monoclinic metal. The ring of disconnected material in the center was due to film damage upon high-current operation.

In conclusion, we probed the transition in a temperature controlled thin film of $VO_2$ using electrical resistance, infrared, visible and x-ray transmission and found that the measurements were considerably different from one another. We further mapped the same film using spatially resolved x-ray mapping during the transition to understand the origins of the differences between the measurements. We observed that the infrared and visible light transmission undergoes a transition at lower temperatures compared to the transition in resistance, during both the heating and cooling, and also that the electronic transition precedes the structural transition during both the heating and cooling. We emphasize the need to employ techniques with high spatial and spectral resolutions with a good control over ambient conditions to study a non-trivial phase transition such as that of $VO_2$.

X-ray measurements were performed at the Advanced Light source (ALS) on beamlines 5.3.2.2 and 11.0.2. The ALS is supported by the Director, Office of Science, Office of Basic Energy Sciences, of the U.S. Department of Energy under Contract No. DE-AC02-05CH11231.